\documentstyle[eqsecnum,aps,epsfig]{revtex}

\newcommand{\be}{\begin{equation}}    
\newcommand{\ee}{\end{equation}}
\newcommand{\beq}{\begin{eqnarray}}
\newcommand{\eeq}{\end{eqnarray}}
\newcommand{\beqn}{\begin{eqnarray*}}
\newcommand{\eeqn}{\end{eqnarray*}}

\def\op{ \ $ }
\def\cl{$ \ }
\def\nn{\nonumber}

\def\ver{\vskip 12pt}

\def\hh{h^{0}_{\ell m}(t,r)}
\def\hs{h^{1}_{\ell m}(t,r)}

\def\h1r{h^{1}_{\ell m}(t,r)}
\def\ha1rp{h^{1'}_{\ell m}(t,r)}
\def\rr{R(t)}
\def\rd{\dot{R}(t)}
\def\fd{\dot{\Phi}(t)}
\def\rdd{\ddot{R}(t)}
\def\fdd{\ddot{\Phi}(t)}
\def\cc{\cos\Phi(t)}
\def\ss{\sin\Phi(t)}

\begin{document}
\draft

\title{Scattering of particles by neutron stars: Time-evolutions
for axial perturbations}

\author
{V. Ferrari$^{1}$ and K. D. Kokkotas$^2$}

\address
{$^1$ Dipartimento di Fisica ``G.Marconi",
 Universit\` a di Roma ``La Sapienza"\\
and \\
Sezione INFN  ROMA1, p.le A.  Moro
5, I-00185 Roma, Italy\\
$^2$ Department of Physics, Aristotle University of Thessaloniki,
Thessaloniki 54006, Greece}

\date{\today}

\maketitle

\begin{abstract}
The excitation of the axial quasi-normal modes  of
a relativistic star by scattered particles is studied by evolving
the time dependent perturbation equations. This work is the first
step towards the understanding of more complicated perturbative processes,
like the capture or the scattering of particles by rotating stars.
In addition,  it may serve as a test  for the results of the full
nonlinear evolution of binary systems.

\end{abstract}


\section{Introduction}
\ver\ver The gravitational interaction of massive bodies in
general relativity is  associated to the emission of gravitational
signals, the features of which depend on the nature of the
interacting bodies and on the characteristics of their 
orbits. To be studied in detail, very energetic processes like the
coalescence of binary systems need the complex machinery of the
fully non linear theory of gravity coupled to hydrodynamics; this
challenging task  has been undertaken by several scientific groups
in the world
\cite{ooharanakamura},\cite{shapiro},\cite{baumgarte},
\cite{shibatakoji}. Waiting for
the outcome of these simulations,  approximation methods which
allow to get preliminary information on these processes are
extremely  useful. 
An example of the way in which approximate methods shed 
light on much more complex phenomena,  is the capture of test particles by a
black hole (see ref. \cite{Oharakojimanakamura} for a review).
On the assumption that the infalling
mass is much smaller than the black hole mass, so that its effect
can be treated as a perturbation of the black hole geometry, it
was shown  that the gravitational
signal emitted in this process exhibits a ringing tail due to the
excitation of the quasi normal modes of the perturbed black hole.
Years later, a fully relativistic numerical simulation of the
gravitational collapse of a massive star to a black hole
\cite{starkpiran} showed that the signal emitted in this process
also exhibits an exponentially damped sinusoidal tail. 
The experience matured in black hole perturbations suggested that 
this tail appears because
the newly born black hole radiates its residual mechanical 
energy oscillating in its quasi-normal modes.

A further element of interest in the use of perturbation theory
is that the results  will serve as a test
for fully nonlinear calculations \cite{fond}. 
For instance,  the nonlinear evolution of the
head on collision of two black holes \cite{ncsa}  has proved to be in
excellent agreement with the results obtained via a perturbative 
approach. 

The gravitational signals emitted by massive particles orbiting
around rotating and non-rotating black holes have been extensively
investigated  \cite{Oharakojimanakamura}, 
and more recently this study has been extended to the  signals
emitted when small masses in open or circular orbit, interact with
a compact non rotating star \cite{kojima}-\cite{ruoff}. By a
direct integration of the equations describing the perturbation of
the star, it has been shown that both its fluid modes and the
spacetime modes can in principle  be excited,  if the scattered
mass is allowed to get sufficiently close to the compact star.
These results have been obtained by Fourier transforming the
perturbation equations, and solving the associated boundary value
problem, with the condition that the solution is regular at the
center of the star, and reduces to a  pure outgoing wave at radial
infinity.

This approach  is very powerful  when masses are in open or closed
orbits around a non rotating  star; however, it cannot be applied if the
mass falls onto the star, due to the lack
of information on the interaction between the particle and the
stellar fluid after the impact. To investigate this kind of
problems, a time evolution approach seems more appropriate than
the analysis in the frequency domain, since it allows to compute the
emitted signal at least up to  when the particle touches
the surface of the star.
It  appears promising also to describe the  perturbations of rotating stars
excited by small masses. Indeed, in that case the coupling between
the axial and polar perturbations which arises as a consequence of
the dragging of inertial frames (\cite{chandraferrarislow},
\cite{Kojima}), leads to a set of equations that are not simply
reducible to wave equations as in the case of a non rotating star.
Moreover, the time evolution approach has recently been used in the 
literature to study the scattering of wave packets by non rotating 
relativistic stars, (\cite{NAKDK}, \cite{Allen}, \cite{Ruoff}, 
\cite{shapiro2}), where it has been shown to correctly
reproduce the spectra of axial and polar perturbations, 
in a perturbative study of colliding neutron stars \cite{Allen2}
(close limit), and  to provide proper
initial data for stellar perturbations \cite{inidata}.

These considerations motivate the work done in this paper: we
develop a scheme for the numerical evolution of stellar
perturbations excited by a scattered mass. 
In order to allow a direct comparison with previously obtained results,
we integrate the equations describing the axial perturbations of a homogeneous
star with increasing compactness, for different values of the
orbital parameters of the scattered mass, and we choose the
same stellar and orbital parameters as in ref. \cite{tominagasaijomaeda},
 where  the axial equations were integrated in the frequency domain.

It should be mentioned that the time-evolution equations
do not suffer of the annoying problem of the divergence of the source term
when the particle transits through the turning point,
which causes some difficulties  in the integration of the
Regge-Wheeler and of the Zerilli equations  in the frequency
domain.

The plan of the paper is the following. In section II we shall
briefly review the time-dependent version of the equations of the
axial perturbations. In section III  we shall present the results
of the numerical integration. Our energy spectra   will be
compared with those found in  \cite{tominagasaijomaeda} and
\cite{priceandrade} and the differences will be discussed. In the
Appendix we shall show that the low-frequency part of the spectrum
can be traced back to the bremsstrahlung radiation emitted by the 
scattered mass.

\section{The equations describing the perturbed spacetime}

In order to describe the non axisymmetric  perturbations of a star
induced by a scattered particle, we expand the perturbed
metric in tensor spherical harmonics, and choose the Regge-Wheeler
gauge. The perturbed line element has the form:
 \beq
\label{pertmetric} \nn ds^2 &=& e^{2\nu} dt^2 - e^{2\mu_2} dr^2 -
r^2 d\vartheta^2-r^2\sin^2\vartheta d\varphi^2
 \nn \\  &+&
2\sin\theta{\partial Y_{\ell m} \over \partial \theta} \left[ \hh
dt d\varphi +\hs dr d\varphi\right]
 \nn \\
&-&\frac{2}{\sin\vartheta}
\frac{\partial Y_{\ell m}}{\partial\varphi}\left[\hh dt
d\vartheta + \hs dr d\vartheta\right] \eeq
We  consider the perturbations of a non rotating star
with uniform energy density $\epsilon=$const
\cite{chandraferrariaxial}. It should be reminded that homogeneous
stars can exist only if their radius $R$ exceeds 9/8 times the
Schwarzschild radius, or $R/M > 2.25$ (Buchdahl limit).
The source of the perturbed Einstein equations 
is given by  the stress-energy tensor of the particle
moving along a geodesic of the Schwarzschild spacetime \op
(T(t),R(t),\Theta(t),\Phi(t))\cl:
\be
T^{\mu\nu}_{(p)}=m_0 \frac{dT(t)}{d\tau}
\frac{dz^\mu}{dt}\frac{dz^\nu}{dt}
\frac{\delta(r-R(t))}{r^2}\delta(\Omega-\Omega(t)). \ee 
With this
source, and assuming that the particle moves on the equatorial
plane $\Theta=\frac{\pi}{2}$, the axial perturbations can be shown to
be described by the following equation
(\cite{chandraferrariaxial}, \cite{tominagasaijomaeda})
\beq \label{waveq1} \frac{d^2 Z_{\ell
m}(t,r)}{dt^2}&-& \frac{d^2 Z_{\ell m}(t,r)}{dr_*^2}+ V_\ell(r)
Z_{\ell m}(t,r) \nn \\
 &=&-16\pi\imath e^{2\nu}\left[
r\left(e^{2\nu} D_{\ell m} \right)_{,r}-  e^{2\nu} Q_{\ell
m}\right] \eeq where
\begin{equation}
 Z_{\ell m}(t,r)= \frac{e^{\nu-\mu_2}}{r}h_1(t,r)
\end{equation}
is a gauge invariant quantity, $r_{*}$ is the tortoise coordinate
and
\begin{equation}
V_{\ell}=(e^{2\nu}/r^3)\left[\ell(\ell+1)r+r^3(\epsilon-p)-6m(r)\right]
\end{equation}
is the potential, which reduces to the Regge-Wheeler potential
outside the star, where pressure and density vanish
($p=\epsilon=0$). The two functions $D_{\ell m}$ and $Q_{\ell m}$,
related to the particle's motion, have the following form: \beq
\label{Dlm} D_{\ell m}(r,t)&=&-\frac{\mu m
}{\ell(\ell+1)(\ell-1)(\ell+2)} ~\frac{L^2_z}{E}~ \frac{e^{2\nu}}{
r^4}~ \delta(r-R(t)) \cdot  Y^\ast_{\ell m,\vartheta}
\left|_{\Theta=\frac{\pi}{2},\Phi(t)} \right.
\\
\label{Qlm} Q_{\ell m}(r,t)&=& -\frac{\imath \mu
e^{-2\nu}}{\ell(\ell+1)} \frac{L_z}{ r^3} \frac{dR}{dt}~\delta
(r-R(t)) \cdot {Y^\ast_{\ell m,\vartheta}}\left|_{\Theta=
\frac{\pi}{2},\Phi (t)}\right. \eeq where $\mu$ is the mass of the
particle.

The one-sided energy  spectrum of gravitational waves at infinity will be
calculated from the following relation
\begin{equation}
{dE\over d\omega} = {1\over 16 \pi^2} \sum^\infty_{\ell=2}
\sum^\ell_{m=-\ell} (\ell-1)\ell(\ell+1)(\ell+2) |Z_{\ell m}(\omega) |^2,
\label{spectrum}
\end{equation}
where \op Z_{\ell m}(\omega)\cl is the Fourier transform of the
wavefunction \op Z_{\ell m}(t,r),\cl evaluated at  radial infinity.
For the axial perturbation, we have to deal only with
multipoles with $m=l-1, l-3, \dots$ since the multipoles with $m=l, l-2,
\dots$ correspond to polar perturbations. For the purpose of this
paper we will present results only for $l=2$ and $m=1$.

\section{Numerical method and  results.}
The numerical procedure we follow to calculate the energy spectrum 
can be described by the following steps
\begin{itemize}
\item For a specific stellar model and for a given set of values 
of the energy and of the angular momentum of the particle,
$(E,L_z)$,
 we calculate the geodesics i.e. the position of the particle, for a 
discrete number of time steps from $r\to \infty$ up to
 $r=r_t$ ($r_t$ is the turning point of the given open trajectory).
\item For the specific geodesic, we estimate the source term of 
(\ref{waveq1}) for each time and spatial grid point. 
Since there is no back reaction this calculation is done only once 
at the beginning.
\item According to equations (\ref{Dlm})-(\ref{Qlm}) the position 
of the particle at each time step is represented by a delta function,
that we approximate by a narrow Gaussian function
\begin{equation}
\delta(r-R(t)) \approx {\alpha \over {\sqrt{\pi}}}
e^{- \alpha^2 (r-R(t))^2 }
\end{equation}
(this approximation has been also used in \cite{ruoff}).
The code has been tested for various values of $\alpha$ in order to
check the way in which the width of the Gaussian affects the results.
\item Finally, eq. (\ref{waveq1}) is evolved using a second order 
finite-difference scheme. The results have been tested for various grid sizes
in order to achieve the required accuracy.
\end{itemize}

\begin{figure}
\centerline{ \epsfig{figure=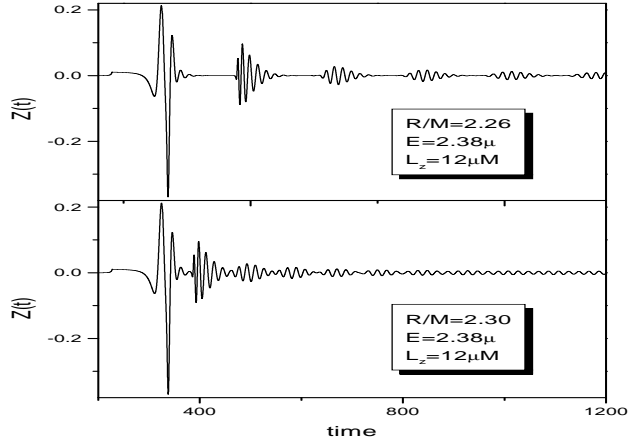,width=9cm,height=7cm} }
\caption{ The waveforms of the signal emitted by a particle with
$(E,L_z)=(2.38\mu,12\mu M)$ scattered by a star with $R/M=2.26$
(upper panel) and $R/M=2.30$ (lower panel).  The excitation of the
{\em w-modes} is clearly manifested in the signal. } \label{fig1}
\end{figure}

\begin{figure}
\centerline{ \epsfig{figure=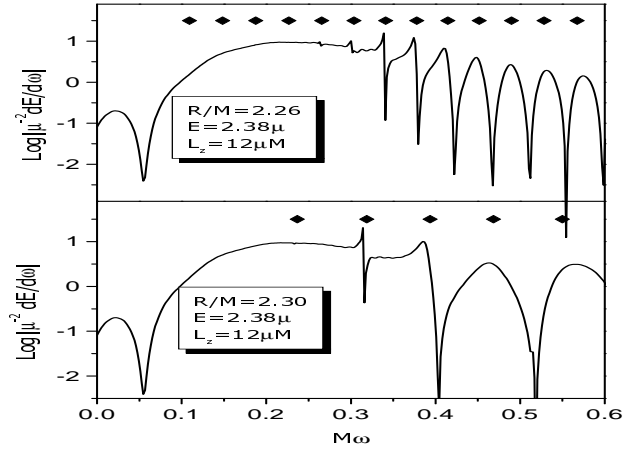,width=9cm,height=7cm}}
\caption{ The energy spectrum of the signal emitted by a particle
with $(E,L_z)=(2.38\mu,12\mu M)$  scattered by a star with
$R/M=2.26$ (upper panel) and $R/M=2.30$ (lower panel). The sharp
peaks correspond to the excitation of the
{\em w-modes}, while the broader
peaks at low frequency are mainly  contributed by the quadrupole radiation 
of the particle due to its orbital motion.
The diamonds correspond to the real part of the known
quasi-normal mode frequencies.
It seems that the lowest frequency trapped
modes of the first model of star are difficult to be excited.
} \label{fig2}
\end{figure}

\begin{figure}
\centerline{ \epsfig{figure=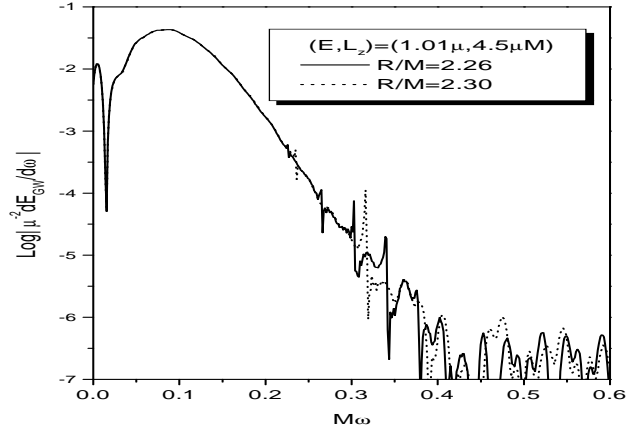,width=9cm,height=7cm}}
\caption{ The energy spectrum of the signal emitted by a particle
with $(E,L_z)=(1.01\mu,4.5\mu M)$,  scattered by a star with
$R/M=2.26$ and  with $R/M=2.30$. The sharp peaks correspond to the
excitation of the various {\em w-modes} of the star,
whereas the dominant contribution to the emitted energy
is that of the quadrupole emission of the scattered
particle, which produces the broader peaks at lower frequencies.}
\label{fig3}
\end{figure}

In Figure \ref{fig1} we plot two waveforms corresponding to 
a particle with orbital parameters 
$(E,L_z)=(2.38\mu,12\mu M)$  scattered by two homogenous stars, one 
with compactness $R/M=2.26$  and the other with $R/M=2.30$.
The initial part of the signal is similar in both cases, since it is 
due  mainly to the  accelerated particle, whereas  the subsequent
oscillations  have different frequencies and damping times
since they are due to the excitation of the axial modes 
of the two stars. The energy spectrum of the above signals, 
calculated via  eq.  (\ref{spectrum}), is shown in Figure 
\ref{fig2}. On the same figure, we mark the location of the real part of
the frequency of the axial quasi-normal modes of the two 
stars, as computed in \cite{kokkotas94} by solving
the homogeneous equation as a boundary value problem.
It  comes out that the lowest trapped modes, that have much longer damping
times, do not seem to be excited,
whereas the higest w-modes are clearly excited in both cases.
In Figure \ref{fig3} we show the energy spectrum for a less energetic
particle $(E,L_z)=(1.01\mu,4.5\mu M)$; here
one can easily identify the peaks corresponding to the
excitation of the w-modes, but the
bulk of the emitted energy is due to the quadrupole emission of the
accelerated particle. This continues to be
true if the central star is less relativistic. These results
suggest that in order to have a significant excitation of the
stellar axial modes  one needs both the periastron of the
scattering orbit to be as close as possible to the stellar surface,
and the star to be of high compactness. The above features
agree with the results of ref. \cite{ferrarigualtieribor}, where
the axial and polar emission for particles scattered by a compact 
polytropic star was studied in the frequency domain, and with those
presented in \cite{ruoff}, where the  polar mode
excitation by scattered particles has been studied with a time-evolution
approach.

Our results also qualitatively agree
with those of ref. \cite{tominagasaijomaeda}, as 
a comparison of our figure 2 and 3 with
their figures 5, 6 or 8 clearly show. 
However, there is a difference of a factor $\sim 2\pi$ 
in the energy spectra, which is probably due to 
a different normalization.
More relevant is the difference we find in the energy spectra
at low frequency:
we observe some peaks that do not appear either in 
the spectra of ref. \cite{tominagasaijomaeda}, or in those presented in
ref. \cite{priceandrade}.
Since this same feature also appeared in previous works
(\cite{ferrarigualtieribor} and \cite{ruoff}), we have 
investigated the low frequency behaviour in some more detail.
We have computed the energy spectrum emitted by the same
scattered  particles in a semirelativistic approximation, by assuming
that the particles move along a geodesic of the background geometry, but
radiate as if they were in flat spacetime, according to the quadrupole
formula. The energy spectra estimated by this procedure are shown in
Figure \ref{fig4} for the star with 
$R/M=2.26$  and  for 
$(E,L_z)=(2.38\mu,12\mu M)$ and $(E,L_z)=(1.01\mu,4.5\mu M)$ 
(details of this calculation are given in the Appendix).
We find  low frequency peaks which correspond to those 
in figures \ref{fig2} and \ref{fig3}; this indicates that the low
frequency part of the energy spectrum is  ``shaped" by the
bremsstrahlung radiation emitted by the accelerated particle.
It should be mentioned that one should not expect
an exact coincidence between the results of the
semirelativistic approximation and of the  relativistic approach, 
especially when the orbit of the scattered particle is very relativistic.
However, the two approaches must tend
to the same limit when the particle moves on less relativistic orbit, and
we have checked that this is indeed the case.

As a concluding remak, we would like to point out that the calculations
presented in this paper  prove that the
numerical evolution of the time dependent perturbation equations
is a fast and accurate way to study scattering processes occurring in the
vicinity of a  neutron stars; they allow to obtain detailed information
either on the amount of energy emitted  at low frequency, 
which can be traced back to the quadrupole emission, 
and on  the excitation of the neutron star quasi-normal modes.
In a future work we plan to extend the present investigation to the study 
of the excitations of the stellar modes of rotating stars. 
There, the coupling of the axial and polar perturbations makes the whole
process more complicated and the numerical evolution of the 
perturbation equations appears a promising approach.
\section*{Acknowledgements}
Helpful discussions with E.Berti and J.Ruoff have been  greatly
appreciated. We are also grateful for a grant from the INFN 
which has enabled our collaboration.

\section*{APPENDIX: The  quadrupole emission of the scattered mass}
In order to understand what is the contribution of the gravitational
radiation emitted by the accelerated particle to the total energy spectrum,
we have used a semi-relativistic approximation introduced in 1971
\cite{ruffiniwheeler}, which assumes that
the particle moves along a geodesic of the curved spacetime, but radiates
as if it were in flat spacetime. On this assumption, using the quadrupole
formula, we have computed the TT-components of the gravitational wave 
emitted by  the scattered mass.
As usual, the reduced quadrupole moment  is
\begin{equation}
Q_{kl}=
\mu \left(X^{k}X^{l}-\frac{1}{3}\delta^{k}_{l}\left|{\mathbf{X}}
\right|^{2}\right),
\end{equation}
where  the two-dimensional vector \op {\mathbf{X}}\cl is the position of the
particle along its trajectory in the equatorial plane $ {\mathbf{X}}= \left(
R(t)\cos\Phi(t), R(t)\sin\Phi(t) \right)$, and 
\op R(t)\cl and \op \Phi(t)\cl are given by the geodesic
equations. 
The expressions of the second time derivative of the
components of \op Q_{kl}\cl in terms of \op R(t)\cl and \op \Phi(t)\cl
are
\beq \label{secondder1}
\frac{1}{\mu} \ddot{Q}_{xx} &=&  \left[2 \rd^2+2 \rr
\rdd-4 \rr^2 \fd^2 \right]\cc^2 \nn \\ &-&\left[ 8 \rr\fd\rd
+2\rr^2 \fdd \right]\ss\cc
 \nn \\
  &-& \frac{2}{3}\left(\rd^2 + \rr \rdd\right) +2 \rr^2\fd^2 \\
  \label{secondder2}
\frac{1}{\mu} \ddot{Q}_{yy}&=&  -\left[2 \rd^2+2 \rr \rdd-4 \rr^2 \fdd
\right]\cc^2\nn \\ &+&\left[ 8 \rr\fd\rd +2 \rr^2 \fdd
\right]\ss\cc \nn \\ &+& \frac{4}{3}\left(\rd^2 + \rr \rdd\right)
-2 \rr^2\fd^2 \\
 \label{secondder3}
\frac{1}{\mu} \ddot{Q}_{xy} &=& \left[ 8 \rr\fd\rd +
2 \rr^2 \fdd \right]\cc^2\nn \\
  &+&\left[2 \rd^2xa+2 \rr \rdd-4 \rr^2 \fd^2
\right]\ss\cc
 \nn \\
 &-& \rr^2\fdd -4 \rr\fd\rd \\
 \label{secondder4}
\frac{1}{\mu}  \ddot{Q}_{zz}&=& - \frac{2}{3}\left(\rd^2 + \rr \rdd\right)
  \eeq
From these expressions  the non vanishing TT-components
of the emitted wave can be computed as follows
\begin{equation}
h_{ij}^{TT}(t,\ {\mathbf{x}})  =  \frac{2}{r} \left(
P_{ik}P_{jl}-\frac{ 1}{2}P_{ij}P_{kl} \right)
\ddot{Q}_{kl}(\tau)|_{\tau=t-\frac{r}{c}}
\end{equation}
where \op P_{ik}=\delta^{i}_{k}-n_{i}n_{k}\cl is the projector
onto the 2-sphere \op r=const,\cl and ${\mathbf{n}}$ is the radial
unit vector. 
The explicit expression of the two independent components is
 \beq \label{hthetatheta}
r h^{TT}_{\vartheta\vartheta} &=&
 \left[\left(\ddot{Q}_{xx}-\ddot{Q}_{yy} \right)\cos^2\varphi +2
\ddot{Q}_{xy} \sin\varphi\cos\varphi
 \right] \left( 1+\cos^2\vartheta\right) \nn \\
&+&\left[\ddot{Q}_{yy}-\ddot{Q}_{zz}  \right]\cos^2\vartheta +
\ddot{Q}_{zz}- \ddot{Q}_{xx}\\
 \label{hphiphi}
r h^{TT}_{\vartheta\varphi}
&=&-2\cos\vartheta \left[ \left(\ddot{Q}_{xx}-\ddot{Q}_{yy}\right)
\sin\varphi\cos\varphi+ \ddot{Q}_{xy}\left(1-2\cos^2\varphi\right)
\right] 
\eeq 
where \op\vartheta,\varphi\cl are the polar angles.
Using the geodesic equations  the \op \ddot{Q}_{ij}(t)\cl given by
eqs. (\ref{secondder1})-(\ref{secondder4}) can be computed. Then,
by Fourier-transforming  the metric  components given in eqs.
(\ref{hthetatheta})-(\ref{hphiphi}) \op
h^{TT}_{\vartheta\vartheta}(\omega,r,\vartheta,\varphi),~
h^{TT}_{\vartheta\vartheta}(\omega,r,\vartheta,\varphi)\cl are
easily evaluated. Since the energy per unit frequency and unit
solid angle is
\begin{equation}
\frac{dE}{d\Omega d\omega} =\frac{\omega^2 r^2}{16\pi^2} \left\{
\vert
h^{TT}_{\vartheta\vartheta}(\omega,r,\vartheta,\varphi)\vert^2+
\vert h^{TT}_{\varphi\varphi}(\omega,r,\vartheta,\varphi)\vert^2
\right\}
\end{equation}
where the frequency \op\omega\cl is restricted to be positive, the
energy spectrum can be computed  by integrating over the solid
angle. It should be pointed out that our convention on the Fourier
transform is \op f(\omega)=\int_{-\infty}^{+\infty}~f(t) e^{i\omega
t}dt.\cl

\begin{figure}
\centerline{ \epsfig{figure=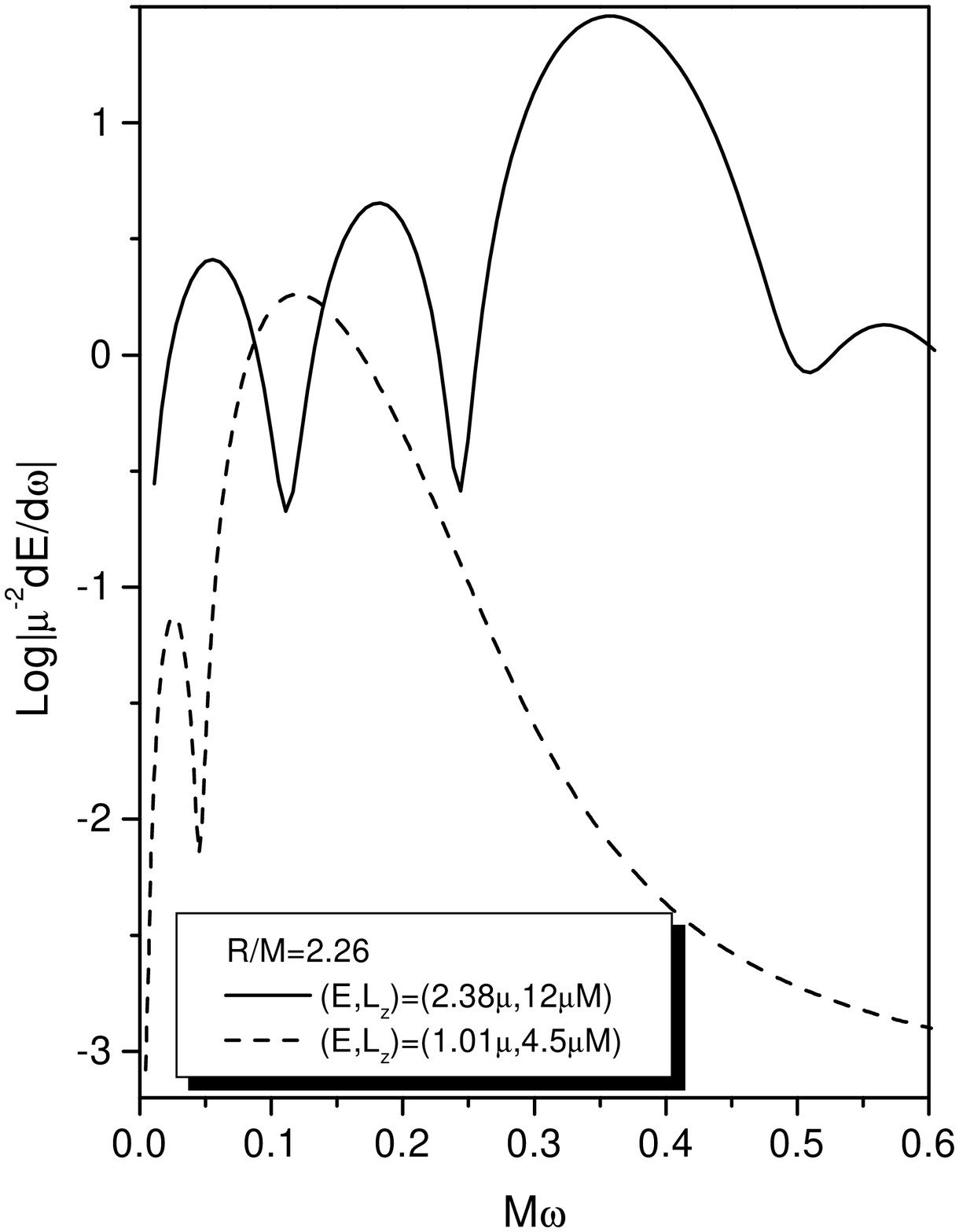,width=9cm,height=7cm}}
\caption{ The energy spectrum computed by the semirelativistic
approximation described in the Appendix, is plotted
for two particles
with $(E,L_z)=(2.38\mu,12\mu M)$, and $(E,L_z)=(1.01\mu,4.5\mu M)$
scattered by  a star with $R/M=2.26$.
A comparison with the energy spectra given in 
figures \ref{fig2} and \ref{fig3}, indicates that the low frequency 
emission is contributed by the the
bremsstrahlung radiation emitted by the accelerated particle.
} \label{fig4} 
\end{figure}

We have numerically evaluated the energy spectrum for the two sets of
values of the orbital parameters considered  in this paper,
i.e. \op \left(L_z=12M, E=2.38\right) \cl and \op
\left(L_z=4.5 M , E=1.01\right),\cl and the results are plotted in
figure \ref{fig4}.

\newpage



\begin{references}
\bibitem{ooharanakamura}
K. Oohara, T.Nakamura,  Prog. Theor.  Phys. Suppl. {\bf 136 }, 270,  (1999)
\bibitem{shapiro}
F.A. Rasio, S.L. Shapiro, Class. Quant. Grav. {\bf 16}, 1, (1999)
\bibitem{baumgarte}
T.W. Baumgarte, S.A. Hughes, S.L. Shapiro, Phys. Rev. D {\bf 60}, 87501, (1999)
\bibitem{shibatakoji}
M. Shibata, K. Uryu, gr-qc/9911058 (1999)

\bibitem{Oharakojimanakamura}
T.Nakamura, K. Oohara, Y. Kojima,  Prog. Theor.  Phys. Suppl. {\bf 90}, 1 (1987)


\bibitem{starkpiran}
R.F.Stark, T.Piran,  Phys. Rev. Lett. {\bf 55}, 891 (1985)

\bibitem{fond}
T.Fond, N.Stergioulas and K.D.Kokkotas MNRAS, 313, 678(2000)

\bibitem{ncsa}
P.Anninos,D.Hobill, E.Seidel, L.Smarr and W.-M.Suen Phys. Rev.
Lett. {\bf 71} 2851 (1993)


\bibitem{kojima}
Y. Kojima,  Prog. Theor. Phys.  {\bf 77}, 297 (1987)

\bibitem{ferrarigualtieribor}
V. Ferrari, L. Gualtieri, and A. Borrelli,  Phys. Rev. D {\bf 59}, 1240 (1999)


\bibitem{tominagasaijomaeda}
K. Tominaga, M. Saijo, K. Maeda,  Phys. Rev. D {\bf 60}, 24004 (1999)

\bibitem{priceandrade}
Z. Andrade, R. H. Price,  Phys. Rev. D {\bf 60}, 104037 (1999)

\bibitem{ruoff}
J.Ruoff, P.Laguna and J.Pullin, gr-qc/0005002 (2000)



\bibitem{chandraferrarislow}
S.Chandrasekhar, V.Ferrari,  Proc. R. Soc. Lond. {\bf A433}, 423
(1991)

\bibitem{Kojima} Y.Kojima, Phys.Rev. D {\bf 46}, 4289 (1992)

\bibitem{NAKDK}
N.Andersson and K.D.Kokkotas Phys. Rev. Lett. {\bf 77},4134 (1996)
\bibitem{Allen}
G.Allen, N.Andersson, K.D.Kokkotas and B.F.Schutz Phys. Rev. D {\bf
58}, 124012 (1998)
\bibitem{Ruoff}
J.Ruoff Phys. Rev. D in press gr-qc/0003088
\bibitem{shapiro2}
V.Pavlidou, K.Tassis, T.W.Baumgarte, S.L.Shapiro Phys. Rev. D in
press gr-qc/0007019
\bibitem{Allen2}
G.Allen, N.Andersson, K.D.Kokkotas, P.Laguna, J.Pullin and J.Ruoff
Phys. Rev. D {\bf 60} 104021 (1999)
\bibitem{inidata}
N.Andersson, K.D.Kokkotas, P.Laguna, Ph.Papadopoulos and
M.S.Shipior Phys. Rev. D {\bf 60} 104021 (1999)

\bibitem{chandraferrariaxial}
S.Chandrasekhar, V.Ferrari,  Proc. R. Soc. Lond. {\bf A432}, 247
(1991)

\bibitem{kokkotas94} K.D.Kokkotas, M.N.R.A.S. {\bf 268}, 1015 (1994);
Erratum {\bf 277}, 1599 (1995)

\bibitem{ruffiniwheeler} R. Ruffini, J. A. Wheeler, {\it Proceedings of
the Cortona Symposium on Weak Interaction} edited by L. Radicati, Roma,
Accademia Nazionale dei Lincei, 169 (1971)
\end{references}
\end{document}